# Hard X-rays as *pump* and *probe* of atomic motion in oxide glasses

Authors:
B. Ruta[1*], F. Zontone[1], Y. Chushkin[1], G. Baldi[2], G. Pintori[2], G. Monaco[2], B. Rufflé[3], and W. Kob[3]
ruta@esrf.fr

Affiliations:
[1)] ESRF- The European Synchrotron, F-38043, Grenoble, France.
B. Ruta, F. Zontone, Y. Chushkin

[2)] Dipartimento di Fisica, Trento University, I-38123 Povo, Trento, Italy.
G. Baldi, G. Pintori, G. Monaco

[3)] Laboratoire Charles Coulomb (L2C), UMR 5221 CNRS-Université de Montpellier, Montpellier, FR-34095, France
B. Rufflé, W. Kob

**Abstract**
Nowadays powerful X-ray sources like synchrotrons and free-electron lasers are considered as ultimate tools for probing microscopic properties in materials. However, the correct interpretation of such experiments requires a good understanding on how the beam affects the properties of the sample, knowledge that is currently lacking for intense X-rays. Here we use X-ray photon correlation spectroscopy to probe static and dynamic properties of oxide and metallic glasses. We find that although the structure does not depend on the flux, strong fluxes do induce a non-trivial microscopic motion in oxide glasses, whereas no such dependence is found for metallic glasses. These results show that high fluxes can alter dynamical properties in hard materials, an effect that needs to be considered in the analysis of X-ray data but which also gives novel possibilities to study materials properties since the beam can not only be used to *probe* the dynamics but also to *pump* it.

**Introduction**
Usually the interaction between X-rays and matter is weak and therefore they are an excellent probe to study the properties of materials[1]. However, the tremendous increase in brilliance of modern X-ray sources such as third-generation synchrotrons and free-electron lasers allows now to use photon fluxes that are so high that one cannot assume any more that the probe beam does not affect the properties of the sample. It is therefore important to obtain a solid understanding on how this strong flux influences the measurements since only this knowledge will allow a correct interpretation of the obtained results.
As it has been documented before, X-rays can in fact induce chemical rearrangements in soft materials like polymers and biological samples[2]. Here the ionizing action of the radiation can modify the structure of the system and even result in its complete disruption. In these cases two main processes have been identified[3,4]. A first primary damage occurs on femtosecond timescales, and is related to the creation of free radicals by photoelectric absorption and Compton scattering[3]. The diffusion of these radicals can then lead to a secondary damage through the formation of additional radicals and the breaking of chemical bonds. As this second process occurs on a timescale of microseconds to milliseconds at room temperature, it can be largely suppressed by working at cryogenic temperatures[5]. Although it is impossible to completely remove the radiation damage, studies on protein crystallography show that it is nevertheless possible to get unique intrinsic properties of a given system[6-8]. For the case of soft materials the radiation damage can limit the achievable resolution[9]. However, this problem can be considerably alleviated by means of fast single shot exposure at high brilliance and fast X-ray sources as free-electron lasers[10,11].

In hard condensed matter X-rays can affect the system through three main mechanisms: Radiolysis, knock-on events and electron rearrangements[12,13]. The first process occurs under UV and X-rays irradiation and leads to the formation of electron-hole pairs and atomic rearrangements. The second effect manifests itself via the displacements of atoms by collision processes that lead to the formation of vacancy-interstitial pairs, such as Frenkel defects[13]. This process usually requires energies as high as 100 keV in order to break atomic bonds in the irradiated materials. The third process are electron rearrangements which occur in presence of pre-existing (natural or induced) defects that serve as electron and hole trap.

Recently Leitner *et al.* investigated for the first time the interaction between hard X-rays and crystalline materials by looking directly at its consequence on the atomic diffusion[14]. That study was done by using X-ray Photon Correlation Spectroscopy (XPCS)[15], a technique which has emerged as a powerful tool to probe the atomic motion in hard crystalline[16] and amorphous materials[17]. By collecting series of diffuse scattering patterns with coherent X-rays, XPCS measures the atomic motion through the temporal intensity fluctuations of the speckles generated by the interference of the waves scattered by the atoms in the material. The detailed analysis of Leitner and co-workers shows that the radiation impinging on crystalline alloys can be treated in the linear response regime, i.e. the perturbation due to the beam is negligible[14].

Here we show that things change completely for the case of simple oxide glasses such as vitreous silica and germania, where the X-rays generate a non-trivial stationary dynamics. This surprising result resembles the one recently observed with transmission electron microscopy[18] and indicates that X-rays do not only probe the atomic motion of the material but can simultaneously also *pump* it. In strong contrast to this behaviour, we find that the intrinsic dynamics of metallic glasses is not affected by the X-rays as in crystalline alloys[14]. Differences in electronic properties and the way electronic excitations couple to phonons are probably the reason for the different behaviour of these two classes of amorphous materials.

**Results**

The effect of hard X-rays on the atomic motion of a given material can be studied by comparing dynamical measurements taken for different sample positions and incoming intensities selected by inserting X-ray attenuators along the beam path. Each attenuator is made of a polished silicon single crystal and leads to a decrease of the X-ray flux, $F$, by a factor $\approx 1/e$ at 8.1 keV. We have measured the dynamics with no attenuators ($F_0 \approx 1 \cdot 10^{11}$ ph/s) and with filters of different thickness: $\approx 80$ μm thick ($F_1 \approx 3 \cdot 10^{10}$ ph/s), $\approx 160$ μm thick ($F_2 \approx 1.2 \cdot 10^{10}$ ph/s), and $\approx 240$ μm thick ($F_3 \approx 3.6 \cdot 10^{9}$ ph/s).

Figure 1a shows the normalized intensity autocorrelation function $g_2(Q,t)$ of $SiO_2$ at T=295 K and for a wave-vector $Q_p=1.5$ Å$^{-1}$ corresponding to the position of the maximum in the structure factor. The function $g_2(Q,t)$ is directly related to the density fluctuations in the material and thus provides information on its relaxation dynamics[15]. Lines in the figure are fits using the Kohlrausch-Williams-Watts function $g_2(Q,t)=1+c \cdot \exp(-2(t/\tau)^\beta)$, where $c$ contains information on the experimental contrast and the nonergodicity factor of the glass, $\beta$ describes the shape of the curve, and $\tau$ is the characteristic decay time[15].

Surprisingly the correlation functions of vitreous silica display a full decorrelation to zero, even at ambient temperature, thus a temperature that is only 20% of the glass transition temperature $T_g$, in agreement with previous results for silicate glasses[19,20]. However, in marked contrast with those works, the decorrelation on the atomic scale observed here cannot be ascribed to *spontaneous* density fluctuations as it depends strongly on the incident flux of the sample in that the decay shifts toward longer times upon decreasing *F*. Since the decay time increases by about two orders of magnitude when the flux is decreased by the same amount we expect that $\tau$ is inversely proportional to *F*. This is confirmed by the presence of a master curve in Fig. 1b where we show the same data as in panel a), but now as a function of the time normalized to the flux. Similar results have been obtained also for $GeO_2$ (see Fig. S1), while this effect is absent in metallic glasses (Fig. 1c), in agreement with the work of Leitner and coworkers[14]. As discussed below, the marked difference in the atomic motion of oxide and metallic glasses can be attributed to the different manner how metals and insulators react on the atomic scale to the deposit of the energy carried by the X-rays and the concomitant *real* occurrence of microscopic structural rearrangements in metallic glasses, as confirmed also by several X-ray diffraction studies[21-23].

The observed dependence of the relaxation dynamics on the incident flux suggests that in the absence of such a flux silica glass is in an arrested state, with relaxation times likely too large to be observed during the experimental time scale considered here. We have tested this hypothesis with a set of measurements that had a constant lagtime, $\Delta t$= 6.15 s, between two consecutive images. This time is the sum of the exposure time $\Delta t_e$ (beam on), the sleeping time $\Delta t_s$ (beam off), and the readout time $\Delta t_r = 1.15$ s (beam off), i.e. $\Delta t = \Delta t_e + \Delta t_s + \Delta t_r$. Figure 2a clearly shows that the measured decay time $\tau$, increases as we reduce the exposure time. This process follows approximately the relationship: $\tau = \tau_0 \frac{\Delta t}{\Delta t_e}$, where $\tau_0$ is the decay time for continuous exposure and negligible readout time ($\Delta t_s = \Delta t_r = 0$). For example, $g_2(Q,t)$ measured with a continuous exposure of $\Delta t_e = 5$ s per frame (red symbols) is approximately one order of magnitude faster than that obtained by using $\Delta t_e = 0.5$ s and a sleeping time $\Delta t_s = 4.5$ s between frames (purple symbols). This result gives thus evidence that during the sleeping time the sample does not relax on the atomic scale, at least on the scale of 2-3 hours and at the room temperature considered here. As a consequence the measured decay time can thus be expected to be completely controlled by the mean photon flux impinging on the sample between two frames. To test this we plot in Fig. 2b the $g_2(Q,t)$ of Fig. 2a as a function of the mean flux $<F> = F_0 \frac{\Delta t_e}{\Delta t}$. The almost perfect overlap of the three curves clearly demonstrates that the measured dynamics is indeed triggered by the X-ray beam acting both as a pump and a probe. Another way to see this is to keep the exposure time fixed and to vary the sleeping time (Fig. 2c). We recognize that the duration of the sleeping time does not affect the relaxation time of the system, thus showing that the latter depends only on the mean flux onto the sample (See Fig. S2 for the same scaling for $GeO_2$).

The results we just discussed imply that the relaxation time $\tau$ of the system is inversely proportional to the mean flux <F>, and thus can be changed basically at will. This is shown in Fig. 2d where we plot $\tau$ for a large set of data taken with different parameters (exposure times, sleeping times, and incident intensities) as a function of the mean flux received by the sample in each frame. Also included is data taken in a second $SiO_2$ sample and in $GeO_2$. It is clear that $\tau$ depends linearly on the inverse of the X-ray flux impinging the sample, suggesting a very slow dynamics in absence of X-rays, at least in these two simple oxide glasses. This result is in agreement with the work of Welch *et al.* who reported for similar systems a relaxation time of about ~27 d at ambient temperature[24].

The effect discussed here cannot be classified as standard radiation damage since the induced dynamics is independent on the accumulated dose deposited on a particular sample position (see S.I.). Furthermore, we have found that the decay time can be reversibly modified by simply changing the intensity of the beam. This is demonstrated in Fig. 3a that shows the two-time correlation function (TTCF) measured in $SiO_2$ while changing the beam attenuation without stopping the measurement. As explained in Ref.[25], each point of the TTCF corresponds to the product of two images acquired at two different times. The effect of the different attenuators is signalled by the abrupt changes in the TTCF profile along the main diagonal whose width is proportional to the decay time of $g_2(Q,t)$. The higher the incoming intensity, the faster is the dynamics and thus the thinner is the intensity broadening along the diagonal. Interestingly, the decay time changes quickly (in less than one frame, thus ≈6 s) as we insert the corresponding attenuator during data acquisition. In addition, for a fixed incoming intensity the profile of the TTCF remains constant with time thus implying that the corresponding dynamics is stationary and does not depend on the total accumulated dose. By averaging the set of images corresponding to each attenuator we obtain five $g_2(Q,t)$ whose decay times and shape parameters are reported in Fig. 3b and 3c. While $\tau$ jumps in a reversible way between the values associated with each flux, $\beta$ remains constant with $\beta=1.38\pm0.09$, which strongly differs from the stretched exponential decay (i.e. with $\beta<1$) observed in silicates[19,20]. Similar compressed decays (i.e. with $\beta>1$) have been reported for soft materials and metallic glasses and they could be associated to a strain field in the material generated by a random distribution of slowly-evolving sources of internal stresses[26,27].

In view of the presented results one might wonder whether the same X-ray induced dynamics occurs also in previously measured silicates[19,20]. Unfortunately, insufficient statistics does not allow to perform the same test as in $SiO_2$. Even if present, the observed effect can be expected to disappear at high temperatures as the relaxation times measured in the glass transition region of sodium tetrasilicate glass (NS4) gently decrease with increasing temperatures and are found to match macroscopic measurements in the supercooled liquid[19]. It is important to note that despite the flux dependence of the relaxation time, the measured dynamics still *provides* physical information on the probed material. The shape of the correlation functions is compressed in the two simple oxide glasses and in the metallic glass investigated here while stretched in silicates, hence independently of whether the dynamics is induced by the X-ray beam or not.

The distinct nature of the probed atomic motion in silica and silicates leads also to a different dependence of the decay time on the wave-vector Q. For $SiO_2$ and $GeO_2$, $\tau$ continuously decreases on increasing Q (Fig. 4 for $SiO_2$ and Fig. S3 for $GeO_2$), whereas it displays an oscillatory behaviour in NS4[19] and in the lead silicates studied by Ross and coworkers[20]. It is important to stress that the *incident flux* impinging on the samples is *the same for* both glasses ($SiO_2$ and NS4) and *all wave-vectors*. Therefore, the intensity cannot be responsible for the observed differences that should instead be ascribed to the details of the local atomic surroundings. In this scenario, the weak increase in $\tau$ observed in Fig. 4 in correspondence to the first maximum of the static structure factor ($Q_p\sim1.5\text{Å}^{-1}$) can be interpreted as an indication of the de-Gennes narrowing typically observed in liquids[28]. Conversely, at low Qs $\tau(Q)$ displays the typical increases observed in other glass-formers liquids[29,30]. What is remarkable, however, is the fact that Fig. 4 shows that the flux-induced relaxation dynamics is independent of Q, which indicates that the associated microscopic process is acting is at work on several length scales.

**Discussion**s
The intriguing findings described above suggest the existence of a dynamic process in vitreous $SiO_2$ and $GeO_2$ whose main features can be summarized as follows:
i) The incident X-ray flux induces an atomic motion at temperatures well below $T_g$. The time scale for this motion is inversely proportional to the photon flux. No decorrelation is observed if the system is not irradiated (at least on the scale of hours).
ii) At fixed flux, the dynamics remains *stationary* and is independent of the accumulated dose.

iii) The decay time depends in a *reversible* way and *almost instantaneously* on the incident flux.
iv) The shape of the correlation functions is independent of the flux and should therefore reflect an intrinsic property of the glass.
v) The flux dependence of the induced dynamics is independent of the wave-vector, at least for intermediate and small Q.
vi) The described induced dynamics is observed for simple oxide glasses. Different glasses display distinct behaviours, in that, for instance, metallic glasses are *not affected* by the X-rays.

The above observations point to a complex beam-activated process which differs from the classical radiation damage reported in XPCS studies on soft materials[31,32]. In these cases, the dynamics clearly varies with the global dose, resulting often in pronounced aging phenomena that are usually triggered by significant structural damage, although there are cases where clear structural changes seem to be absent. The latter situation seems to imply the existence of a threshold for radiation damage that is lower for dynamical studies than the one for structural investigations, in agreement with recent studies for protein diffusion[33].
In our work, the probed glasses show only a weak, almost negligible, structural change within the *global* irradiation dose used for the measurements (less than $\approx 10^4$ s at $F_0$), see Supplemental Information, while we do observe remarkable modifications for larger accumulated dose. For this large dose the X-rays modify the average local structure leading to a decrease of the intensity of the maximum in the static profile and a concomitant increase at $\approx 0.9 \text{Å}^{-1}$ (see Figs. S4 and S5). Despite the occurrence of strong radiation damage at larger accumulated doses, we do not associate the observed dynamics to that damage: If it were the case, the dynamics would evolve with the accumulated dose (and thus the decay time as well) and would not be reversible when the flux is changed (see also Figs. S6 and S7 for further confirmation).
Instead we believe that the effect reported in this work is likely due to *radiolysis*. Here the interaction with the X-rays generates localized electronic excitations with energies well below those necessary for knock-on events but large enough for atomic displacements[4,34]. In order to convert these excitations into a mechanical response, they should have a lifetime of $\approx 1$ ps and couple to the phonons. This explains why metallic glasses do not show the discovered dependence on the flux since their electronic excitations delocalize faster, *i.e.* on the fs timescale. In alloys the atomic motion can be induced only through a direct transfer of momentum and energy by knock-on processes which require energies much higher than those employed here[12,17]. In addition we point out that the glassy state of amorphous alloys cannot be considered *arrested* as in oxide glasses: It is indeed well known from diffraction studies that metallic glasses display structural atomic changes well below the glass transition temperature[21-23]. In this case XPCS directly probes the effect of these changes on the atomic motion[21] which can lead to very *complex dynamical patterns under the same irradiated conditions*[35,36]. We do expect, however, that the flux induced dynamics found in the present work is likely to occur also in other non-conducting systems like polymeric compounds and molecular glasses, i.e. that this is a phenomenon that is not just a particularity of the simple oxide glasses considered here.
The induced dynamics in the simple oxide glasses strongly resembles the one observed in electron transmission microscopy[18] and hints that the X-rays should not only be considered as a spectator but also as an important actor in the probed dynamics. Depending on the competition between the intrinsic and the induced dynamics, this effect can certainly become an issue for the determination of the associated time scales, as in the case presented here. However one should keep in mind that it can be also considered as a *great opportunity to*

*probe physical properties* of materials which cannot be achieved by means of any other technique.

Finally, it is worth highlighting that these beam-induced effects will become increasingly relevant for experiments at next generation synchrotrons[37,38] and at free-electron lasers sources[39]. In the latter case, techniques using high intense coherent beams, such as XPCS, are based on fs pulses that have intensities as high as current synchrotron sources provide in one second[40-45]. It is therefore likely that the above effects will be greatly amplified by non-linear responses of the system to collective excitations induced by the absorption of very short and intense X-ray pulses.

**Methods**

**Sample preparation.** Samples of the oxide glasses were prepared in the form of disks of 5 mm diameter by cutting bulk material with a diamond drill bit. The disks were mechanically polished to a thickness of 50 μm for $SiO_2$ and of 20 μm for $GeO_2$, in order to get the best compromise between the scattering signal and the speckle contrast. Bulk $SiO_2$ was a commercial grade Spectrosil ($SiO_2$-bis from Suprasil F300 with less than 1 ppm OH), while $GeO_2$ was prepared by the usual melt-quenching procedure. $Zr_{65}Cu_{27.5}Al_{7.5}$ metallic glasses were prepared by melt spinning in the University of Göttingen. The resulting ribbons had a thickness of about 40 μm.

**XPCS measurements.** We performed several XPCS experiments at the beamline ID10 at ESRF by using 8.1 KeV radiation produced by three undulator sources. The coherent part of the beam (8 μm x 10 μm VxH full width half maximum) was selected by rollerblade slits placed upstream of the sample. The incoming intensity at the sample position was monitored continuously with a scintillation detector counting the photons scattered by air. The absolute incident flux is estimated by normalizing to scattered intensities from kapton foils[46]. The samples were inserted in a homemade resistively heated furnace mounted on a diffractometer in horizontal scattering geometry. The coherently scattered photons (speckles) were recorded by an Andor CCD device installed at ≈70 cm from the sample on a detector arm that was rotating around the sample to cover the Q–range 0.3-4 Å$^{-1}$. Correlation functions were obtained following the analysis described in Ref.[47]. Oxides glasses were measured at room temperature while $Cu_{65}Zr_{27.5}Al_{7.5}$ was annealed at T=413 K (thus at $T/T_g$=0.59). The choice of this temperature was dictated by the requirement to place the system in the temporal window where the decay time remains constant and sufficiently fast with $\tau \leq 10^3$ s in standard working conditions at maximum flux (8.1 keV, no attenuators). This situation can be achieved by working at high temperature in the glassy state, where metallic glasses display stationary dynamics, likely related to an intermittent mechanism of aging[35]. At lower temperatures the dynamics is dominated by the typical fast aging of rapidly quenched metallic glasses[17,21] making the test extremely challenging as it would be difficult to disentangle X-rays induced effects and spontaneous changes of the decay time related to aging. For all samples, the measured contrast was ≈2-5% depending on the experimental conditions.

**Acknowledgements.** Prof. K. Samwer and the University of Göttingen are gratefully acknowledged for the metallic glass. We acknowledge H. Vitoux, K. L'Hoste and L. Claustre for the technical support during the XPCS experiments, and Del Maschio for the sample preparation. We gratefully thank A. Madsen and M. Di Michiel for fruitful discussions.

**Author Contributions.** B.R. conceived the project. B.R., F.Z., Y.C., G.B., G.P., G.M. and B. Rufflé performed the experiments; Y.C. provided the programs for the XPCS data analysis; B.R. analysed the data. G.B. and B. Rufflé provided the samples; All authors discussed the results. B.R. wrote the manuscript with contributions from all authors.

**Author Information.** Correspondence and request for the materials should be addressed to B.R. (ruta@esrf.fr)
**Competing financial interests.** The authors declare no competing financial interests.

Figure 1

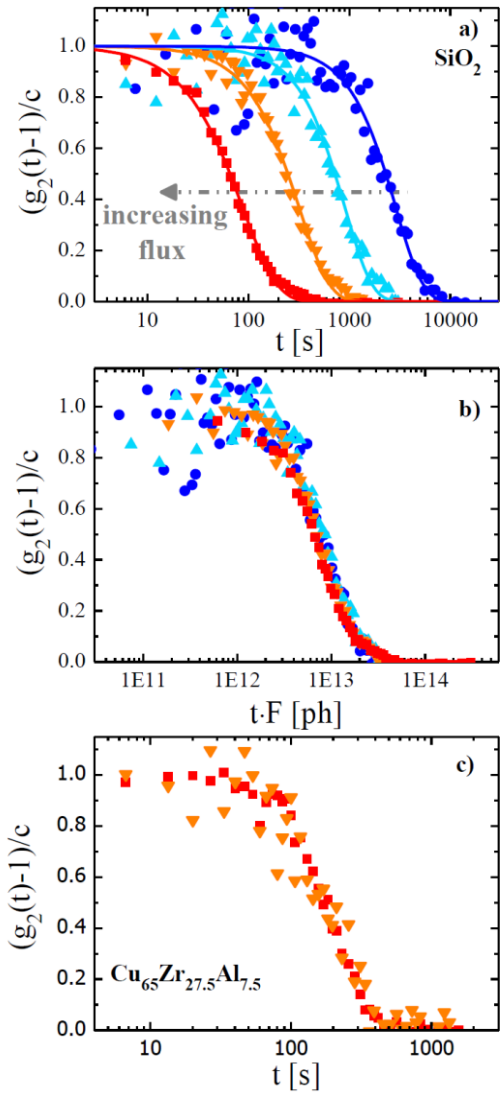

**Figure 1: Relaxation dynamics of the atoms as a function of the incident X-ray beam intensity.** (a) Normalized intensity auto-correlation functions measured in vitreous silica at T=295 K and wave-vector $Q_p=1.5$ Å$^{-1}$ for different intensities of the flux of the X-ray beam. From left to right: $F_0 \approx 1 \cdot 10^{11}$ ph/s (red squares), $F_1 \approx 3 \cdot 10^{10}$ ph/s (orange down-triangles), $F_2 \approx 1.2 \cdot 10^{10}$ ph/s (cyan up-triangles) and $F_3 \approx 3.6 \cdot 10^{9}$ ph/s (blue circles). Lines are fits with a Kohlrausch-Williams-Watts function. (b) Same data rescaled by the incoming flux. (c) Normalized intensity auto-correlation functions measured in $Cu_{65}Zr_{27.5}Al_{7.5}$ metallic glass at T=413 K and $Q_p=2.5$Å$^{-1}$ for $F_0 \approx 1 \cdot 10^{11}$ ph/s (red squares), and $F_1 \approx 3 \cdot 10^{10}$ ph/s (orange down-triangles).

**Figure 2**

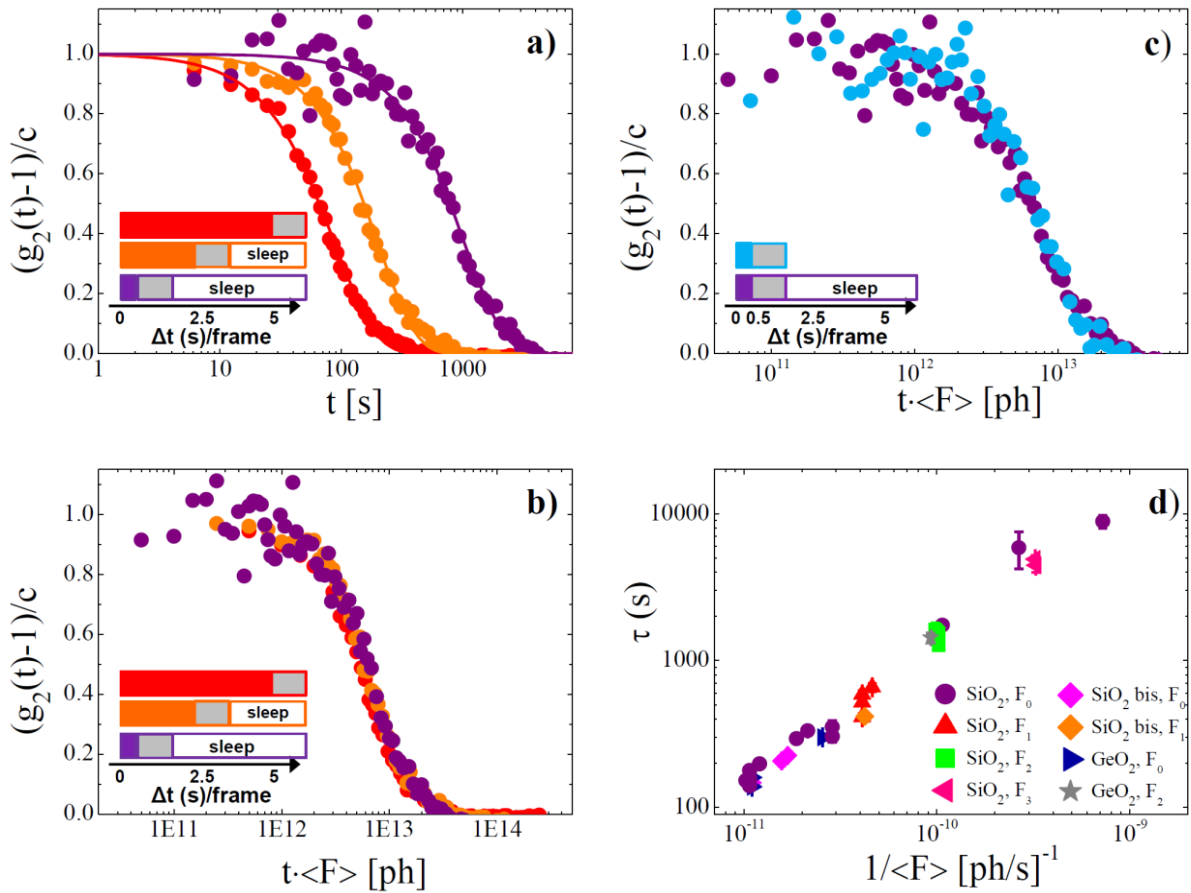

**Figure 2: Tuning of the atomic motion.** (**a**) Normalized intensity auto-correlation functions measured in vitreous silica at T=295 K and $Q_p$=1.5 Å$^{-1}$ with fixed lagtime per frame, $\Delta t$ = 6.15 s, but with different sleeping times $\Delta t_s$ and exposure times $\Delta t_e$. From left to right: $\Delta t_e$=5 s and $\Delta t_s$=0 s (red), $\Delta t_e$=2.5 s and $\Delta t_s$=2.5 s (orange), $\Delta t_e$=0.5 s and $\Delta t_s$=4.5 s (purple). The legend illustrates the acquisition mode per frame with full coloured boxes for the exposure times $\Delta t_e$ (beam on), empty boxes both for the sleeping times $\Delta t_s$ (beam off) and the constant readout time of the CCD $\Delta t_r$ (beam off, grey boxes). (**b**) Same data as in panel (**a**) but now normalized by the mean flux <F>=$F_0 \cdot \Delta t_e / \Delta t$. (**c**) $(g_2(t)-1)/c$ measured with $\Delta t_e$=0.5 s and $\Delta t_s$=4.5 s (same purple data as in panel a) reported as a function of the time t times the mean flux. The data are compared with the $g_2(t)$ measured with the same exposure time $\Delta t_e$=0.5 s and no sleeping time ($\Delta t_s$=0 s, cyan). (**d**) Decay time for different combinations of the incident flux and the lagtime. The data are shown as a function of the inverse mean flux. Also included is data for a second SiO$_2$ sample (SiO$_2$ bis) and for vitreous GeO$_2$. The legend indicates the flux used for each set of measurements.

Figure 3

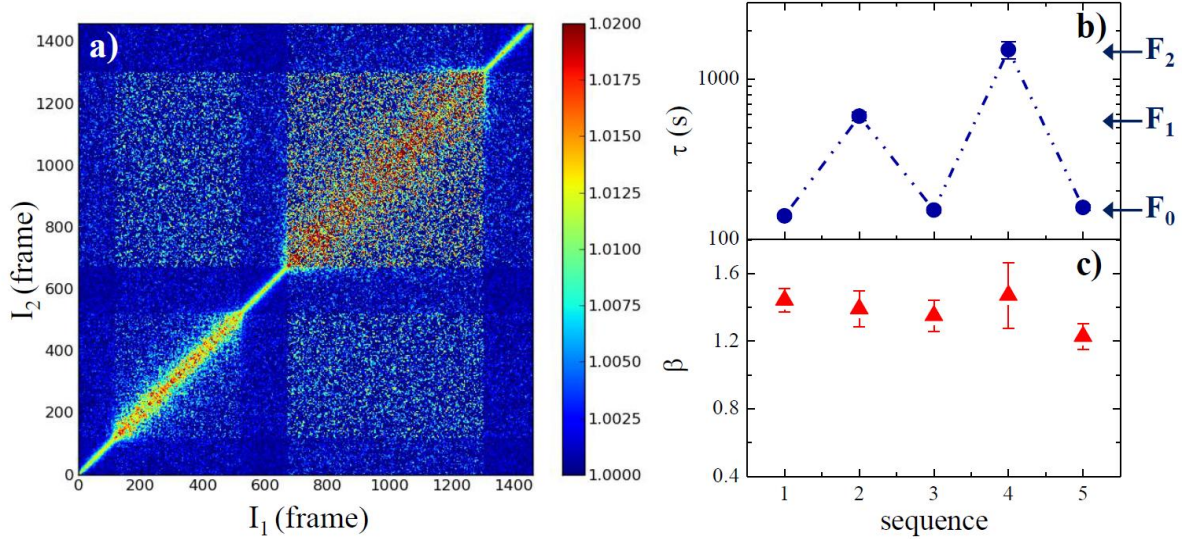

**Figure 3: Instantaneous, reversible and stationary dynamics.** (a) Two-time correlation function measured in vitreous silica at T=295 K and $Q_p$=1.5 Å$^{-1}$ by varying the intensity of the incoming flux. Left to right in frame number: $F_0 \approx 1\cdot 10^{11}$ ph/s, $F_1 \approx 3\cdot 10^{10}$ ph/s, $F_0 \approx 1\cdot 10^{11}$ ph/s, $F_2 \approx 1.2\cdot 10^{10}$ ph/s, and $F_0 \approx 1\cdot 10^{11}$ ph/s. Each frame corresponds to $\Delta t$=6.15 s. (b) Characteristic decay times $\tau$ as a function of the flux intensities used in panel (**a**). (c) Shape parameters $\beta$ as a function of the flux intensities used in panel (**a**).

Figure 4

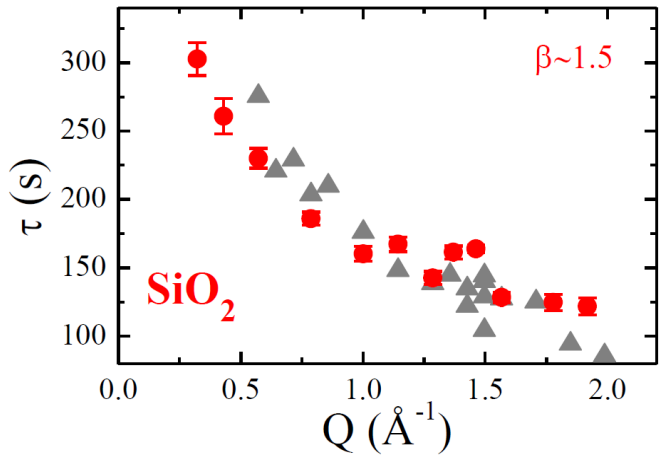

**Figure 4: Wave-vector dependence of the X-ray induced dynamics.** Wave-vector dependence of the characteristic decay time in vitreous silica measured at T= 295K and for $F_0 \approx 1\cdot 10^{11}$ ph/s (red circles). The grey triangles are taken with $F_1 \approx 3\cdot 10^{10}$ ph/s and rescaled by the factor 2.74 corresponding to the X-ray intensity difference between the two measurements.

# SUPPLEMENTARY INFORMATION

## Hard X-rays as *pump* and *probe* of the atomic motion in oxide glasses

Authors:
B. Ruta[1*], F. Zontone[1], Y. Chushkin[1], G. Baldi[2], G. Pintori[2], G. Monaco[2], B. Rufflé[3], and W. Kob[3]

[1] ESRF- The European Synchrotron, F-38043, Grenoble, France.
[2] Dipartimento di Fisica, Trento University, I-38123 Povo, Trento, Italy.
[3] Université de Montpellier and CNRS, Laboratoire Charles Coulomb, UMR 5221, F-34095, Montpellier, France
CNRS, Laboratoire Charles Coulomb UMR 5221, F-34095, Montpellier, France.

## 1. X-ray induced atomic motion in GeO$_2$ at room temperature

Figure S1a shows the intensity auto-correlation function measured in vitreous germania for different values of the incident flux $F$. As it is the case for SiO$_2$, the decay time shifts toward faster time scales if the flux is increased. If the data is plotted as a function of time $t$ times the mean flux, the different data sets superimpose (Fig. S1b).

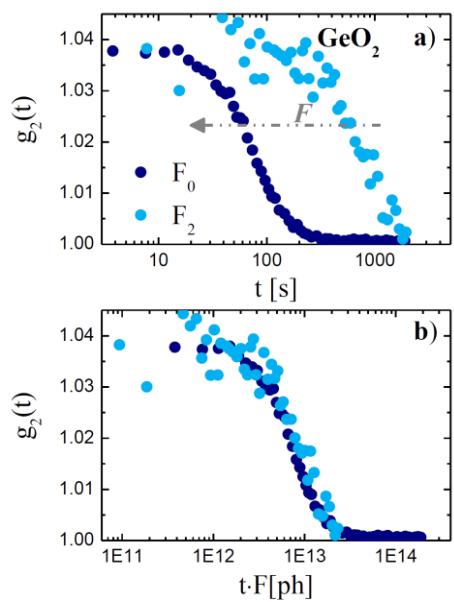

**Figure S1: X-ray induced dynamics in vitreous GeO$_2$.** (**a**) Intensity auto-correlation function measured in vitreous germania at T=295 K and $Q_p$=1.5Å$^{-1}$. Left curve (blue): $F_0 \approx 1 \cdot 10^{11}$ ph/s. Right curve (cyan): $F_2 \approx 1.2 \cdot 10^{10}$ ph/s. (**b**) Same data rescaled for the incoming flux.

This result implies that the decay time is completely fixed by the exposure time and that data taken with the same exposure time $\Delta t_e$ but different sleeping time $\Delta t_s$ perfectly overlap when rescaled for the mean flux impinging on the sample (Fig. S2).

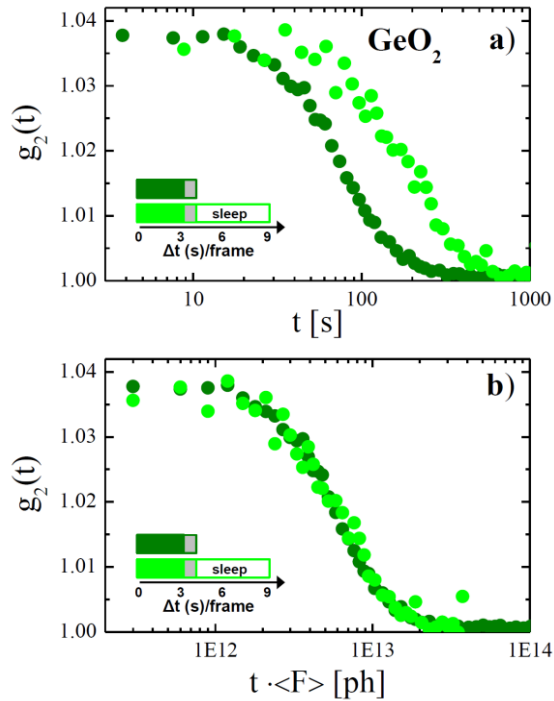

**Figure S2: Characterization of the induced motion.** (a) Intensity auto-correlation functions measured in vitreous germania at T=295 K and $Q_p=1.5$ Å$^{-1}$ with fixed exposure time per frame, $\Delta t_e = 3$ s, without (dark green) and with (light green) sleeping time $\Delta t_s=5$ s between frames. (b) Same data as a function of the time times the mean flux $<F>=F_0 \cdot \Delta t_e / \Delta t$.

Despite the presence of this flux-dependence of the relaxation time, the decay time does reflect the intrinsic properties of the system and varies with the probed wave vector Q, displaying an increase at low Qs as it is the case in SiO$_2$ (Fig. S3).

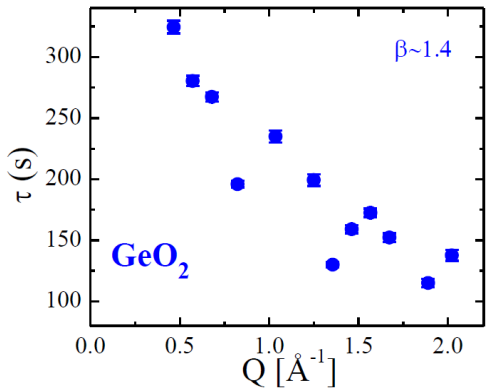

**Figure S3: Wave-vector dependence of the atomic motion.** Wave vector dependence of the characteristic decay time in vitreous germania measured at T= 295K and for $F_0 \approx 1 \cdot 10^{11}$ ph/s (blue circles).

## 2. Effect of the X-rays on the structure and reproducibility of the data

By irradiating always the same spot we do observe eventually a structural change in SiO$_2$. The resulting damage occurs slowly and it is basically negligible for short global irradiated times, i.e. the times considered in the present work, while it becomes significant after $\approx 10^4$ s of irradiation with maximum flux $F_0$ at 8 keV. For larger irradiation times, there is a drop of the intensity at the first maximum of the static structure factor S(Q) which is accompanied by an increase at wave-vectors around Q=0.6 Å$^{-1}$. This is shown in Fig. S4 where we report the intensity static profile measured for different global accumulated dose.

The fact that the damage is very slow at the beginning explains while we do observe stationary and reversible dynamics during the measurements reported in the main manuscript.

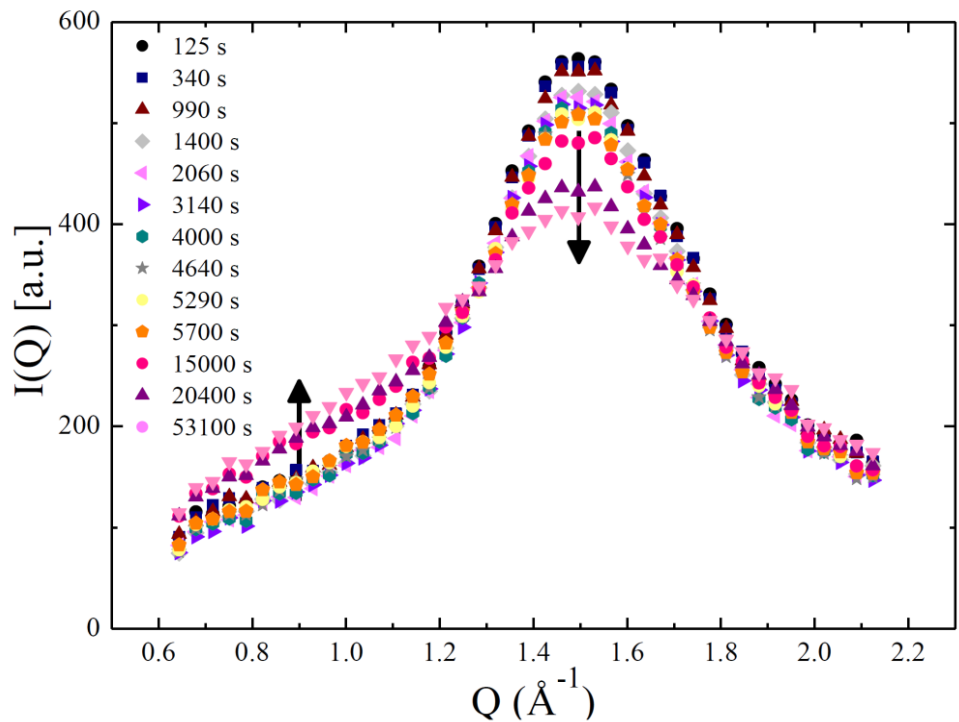

**Figure S4: Effect of the X-rays on the structure.** Static intensity profile measured in vitreous silica at T=295 K for different global irradiated times. The data are normalized by the incoming intensity after background subtraction. The arrows indicate the loss of intensity at the first maximum and the concomitant increase at around 0.9 Å$^{-1}$ due to the X-ray irradiation.

In order to characterize this radiation damage effect quantitatively we have fitted the intensity profile with a Lorentzian and in Fig. S5 we show the so determined fit parameters obtained from samples with different irradiation conditions. Purple circles are taken by irradiating always the same spot with $F_0$. Orange squares are measured always on the same spot with $F_1$, while magenta stars are measured with $F_1$ on different spots which were previously irradiated to measure the dynamics reported in Fig. 5 of the main manuscript (Q dependence measured with $F_1$). From this figure it is clear that the irreversible damage becomes significant only for global irradiation dose higher than $10^4$ s at maximum flux $F_0$.

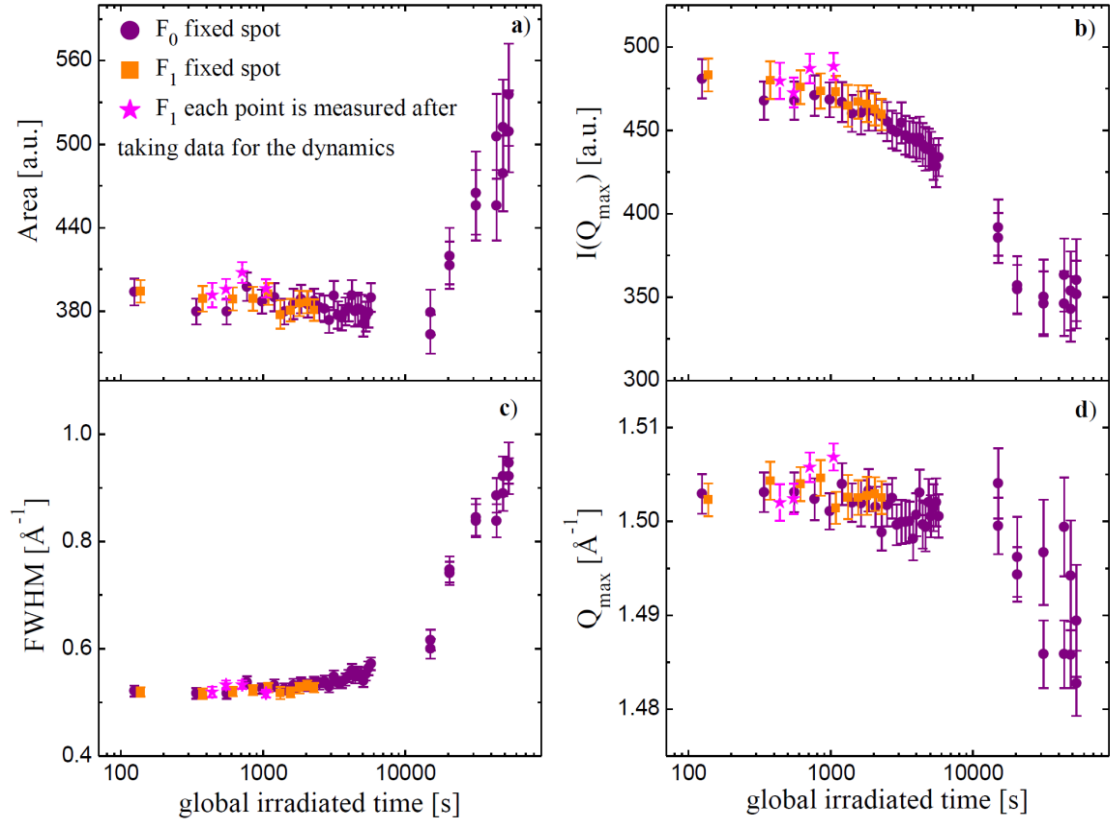

**Figure S5: Characterization of the structural damage.** Integrated area (**a**), intensity of the maximum (**b**), full width at half maximum (**c**), and position of the maximum (**d**) of the static intensity profiles measured in vitreous silica at T=295 K and as a function of the global irradiated times. Purple circles are data taken with $F_0$ on a fixed sample position. Orange squares are measured with $F_1$ always on the same sample position, while magenta starts have been collected with $F_1$ after taking the dynamical data reported in Fig. 5 in the manuscript.

The absence of a signature of the structural damage in our XPCS dynamical data is confirmed by the reversibility and the perfect scaling in the data reported in the main text and is also shown in Fig. S6 where we do report on the left the two-time correlation function measured during an acquisition with maximum flux. The broadening of the intensity along the main diagonal remains constant during the X-rays illumination demonstrating that no damages are occurring in the material. This is further confirmed by the fact that the number of photons collected by the detector follows exactly the behaviour of the incoming intensity (panel b).

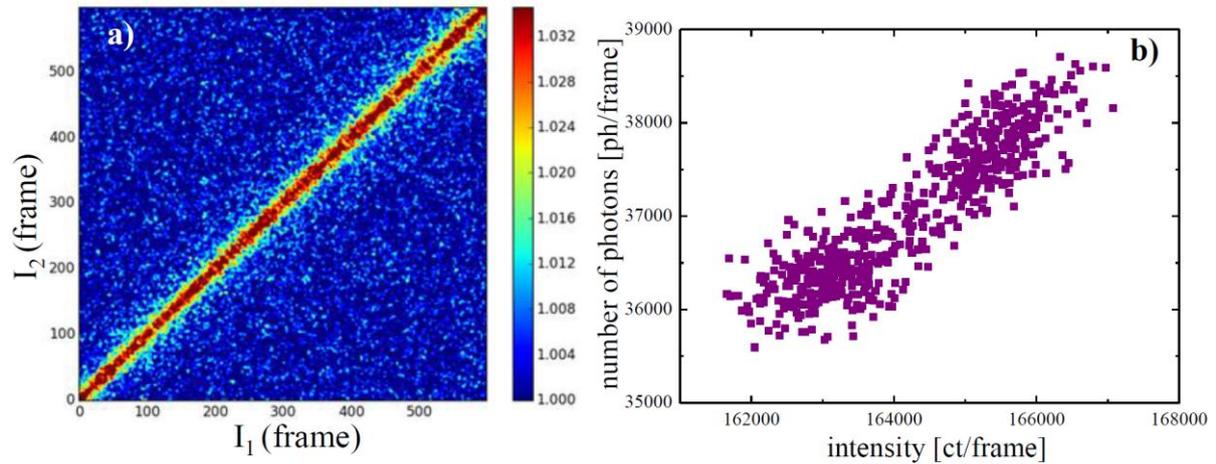

**Figure S6: Absence of structural damage and stationary dynamics for low global accumulated dose.** (**a**) Two-time intensity correlation function measured in $SiO_2$ at T=295 K and $Q_p=1.5 Å^{-1}$ with full intensity $F_0$. (**b**) Corresponding number of scattered photons impinging in the detector as a function of the incoming intensity per frame. The lack of beam damage is confirmed by the straight correlation between these two quantities.

Figure S7 shows the reproducibility of the data with global irradiated dose up to $\approx 10^4$ s. This confirms the independence of the data reported in the manuscript with respect to the global dose and thus the absence of structural damage on this time scale.

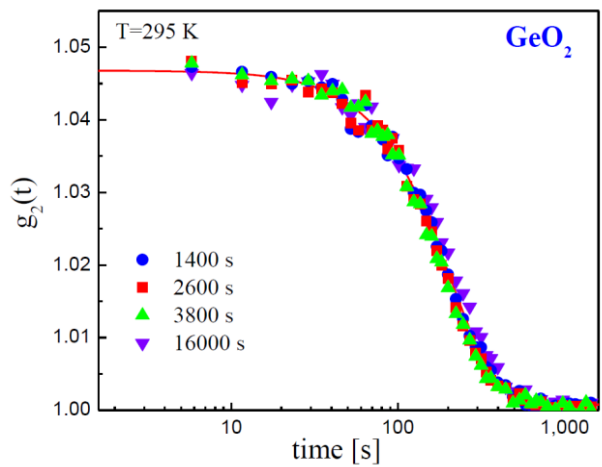

**Figure S7: Reproducibility of the data for low global accumulated dose.** Correlation functions measured in vitreous germania at T= 295K and for maximum flux $F_0 \approx 1 \cdot 10^{11}$ ph/s. The data are taken by irradiating always the same spot with increasing global dose. The dynamics remains stationary within the explored irradiation time.